# How many citations are there in the *Data Citation Index*?[1]

Daniel Torres-Salinas[*], Evaristo Jiménez-Contreras[**] and Nicolas Robinson-García[**]

[*]*torressalinas@gmail.com*
EC3Metrics, EC3 Evaluación de la Ciencia y de la Documentación Científica, Universidad de Navarra, C/ Alhóndiga, 6 2°A, Granada, 18002 (Spain)

[**]*evaristo@ugr.es; elrobin@ugr.es*
EC3 Evaluación de la Ciencia y de la Documentación Científica, Universidad de Granada, Colegio Máximo de Cartuja, s/n, Granada, 18071 (Spain)

**Introduction**
Lately we have witnessed a renewed interest for data sharing and the development of reproducible research (Anon, 2008). Although the claim for transparency in research is not new (King, 1995), in the last few years researchers have been challenged with the management and processing of huge amounts of datasets for conducting large-scale studies in what is known as the 'Big Data' phenomenon (Lynch, 2008). But data sharing practices are relatively common in some fields such as Genomics or Astronomy (Borgman, 2012). Their experience has allowed the development of infrastructure and a slow expansion towards the rest of fields, but still these practices are far from common. In order to promote data sharing practices, journals and evaluation agencies have started to introduce policies that encourage and in some cases, demand authors to share their datasets (an overview of such policies is offered by Borgman, 2012; Torres-Salinas, Robinson-Garcia & Cabezas Clavijo, 2013).

One of the main concerns researchers have for sharing data has to do with the idea that such practices are not 'worth it' as they are time-consuming and are not acknowledged by colleagues and funding bodies. In order to surpass such fear, some authors have analyzed the citation effects of publications sharing data concluding that there is a positive relation between them (Piwowar, Day & Fridsma, 2007; Piwowar & Chapman, 2010). In this context, many tools are being developed in order to track 'impact' of data such as DataCite, CrossRef or Thomson Reuters Data Citation Index (Costas, Meijer, Zahedi & Wouters, 2012). Here we will focus on the latter, a multidisciplinary database launched in 2012 which indexes major data repositories from all areas of the scientific knowledge along with citation data associated to them (Thomson Reuters, 2012).

Torres-Salinas, Martín-Martín and Fuente-Gutiérrez (2014) recently studied the coverage of the Data Citation Index (DCI). From their analysis they concluded that the DCI is heavily biased towards the Hard Sciences, the most common document type is datasets (94% of the total share) and four repositories represent 75% of the database. This paper builds up on their work focusing on the citation distribution of the DCI by areas and by repositories, offering the first citation analysis so far of the DCI.

**Material and methods**
In this paper we conduct an analysis of the citation distribution of the Data Citation Index by areas and repositories. Between May and June, 2013, we retrieved all records indexed in the DCI and created a relational database for data processing. Subject categories to which repositories were assigned were aggregated into four broad areas (Science, Engineering & Technology, Social Sciences and Arts & Humanities). The DCI includes three different document types: datasets, data studies and repositories. However, the distribution of eachof them varies by repository. While some repositories include both datasets and data studies, other only include one of them. Also, not all fields in records seem to be common to all repositories. This can be seen especially in the case of the fields dedicated to assigning keywords to each record.

---
[1] Nicolas Robinson-Garcia is currently supported by a FPU Grant of the Spanish Ministerio de Economía y Competitividad

**Results**

In table 1 we show the main figures by document type. There are a total of 2,626,528 records in the DCI. Most of these are datasets, representing, 94% of the database. Regarding the total number of citation received, 88% of all records remain uncited. Data studies receive more citation in average (0.69) than datasets (0.12), but again, datasets accumulate most of the citation included in the DCI (73%).

Table 1. Indicators for all records, datasets and data studies

|  | All Document Types | Datasets | Data studies |
|---|---|---|---|
| Total Citations | 404,211 | 294,051 | 106,895 |
| Total Records | 2,623,528 | 2,468,736 | 154,674 |
| Uncited Records | 2,311,553 | 2,185,062 | 126,428 |
| % Uncited | 88.11 | 88.51 | 81.74 |
| Citation Average | 0.15 | 0.12 | 0.69 |
| Standard Desviation | 3.06 | 0.36 | 9.56 |

When focusing on the analysis by areas, 81% of the records belong to the area of Science, followed by far by Social Sciences (18%). On the other hand, Engineering & Technology is the most underrepresented area with 0.1% of the whole share. This pattern is also seen when focusing on datasets where Science, was again represents 81% of the database followed by Social Sciences with a share of 17%. However, this picture changes slightly when focusing on datasets. Although the distribution is still severely biased towards Science (74%), Social Sciences has a higher presence (24%). Regarding the citation distribution, only in the area of Engineering & Technology we see a citation average above 0.5, highlighting the high degree of uncitedness. Science accumulates most citations (79%) followed by the Social Sciences (18%), Arts & Humanities (5%) and finally, Engineering & Technology (0.2%). But there are significant differences when analyzing each document type. While in the fields of Engineering & Technology and Science, researchers tend to cite datasets (97% of all citation received in Engineering & Technology and 92% in Science are directed to datasets), the opposite occurs in Social Sciences and Arts & Humanities, where most of the citations were directed to data studies (96% in the case of the former and all except one citation in the case of the latter).

Table 2. Indicators for all records, datasets and data studies by area

A. All document types

|  | Total Records | Total Citations | Citation Average | Standard Deviation |
|---|---|---|---|---|
| Engineering & Technology | 1,786 | 916 | 0.51 | 0.90 |
| Humanities & Arts | 51,444 | 20,460 | 0.40 | 7.99 |
| Science | 2,118,855 | 319,458 | 0.15 | 0.59 |
| Social Sciences | 462,826 | 72,855 | 0.16 | 6.84 |

B. Datasets

|  | Total Records | Total Citations | Citation Average | Standard Deviation |
|---|---|---|---|---|
| Engineering & Technology | 1,545 | 890 | 0.58 | 0.94 |
| Humanities & Arts | 44,588 | 1 | 0.00 | 0.00 |
| Science | 2,004,449 | 293,193 | 0.15 | 0.40 |
| Social Sciences | 424,952 | 7 | 0.00 | 0.01 |

C. Data studies

|  | Total Records | Total Citations | Citation Average | Standard Deviation |
|---|---|---|---|---|
| Engineering & Technology | 240 | 26 | 0.11 | 0.50 |
| Humanities & Arts | 6,847 | 20,459 | 2.99 | 21.72 |
| Science | 114,338 | 26,189 | 0.23 | 1.91 |
| Social Sciences | 37,855 | 69,659 | 1.84 | 17.34 |

This phenomenon is later confirmed when analyzing the citation distribution by subject categories. In figures 1 and 2 we show the top 10 subject categories according to the DCI with a higher number of citations received. Hence, we see that all top ten subject categories for datasets receiving citations belong to the area of Science (Figure 1). Also, we observe that a single subject category, Crystallography, accumulates nearly half of all citations to datasets. Indeed, this category along with Biochemistry & Molecular Biology and Genetics & Heredity represent 86% of all citations.

Figure 1. Top 10 subject categories with a higher number citations received, citation average and standard deviation for datasets indexed in the Data Citation Index.

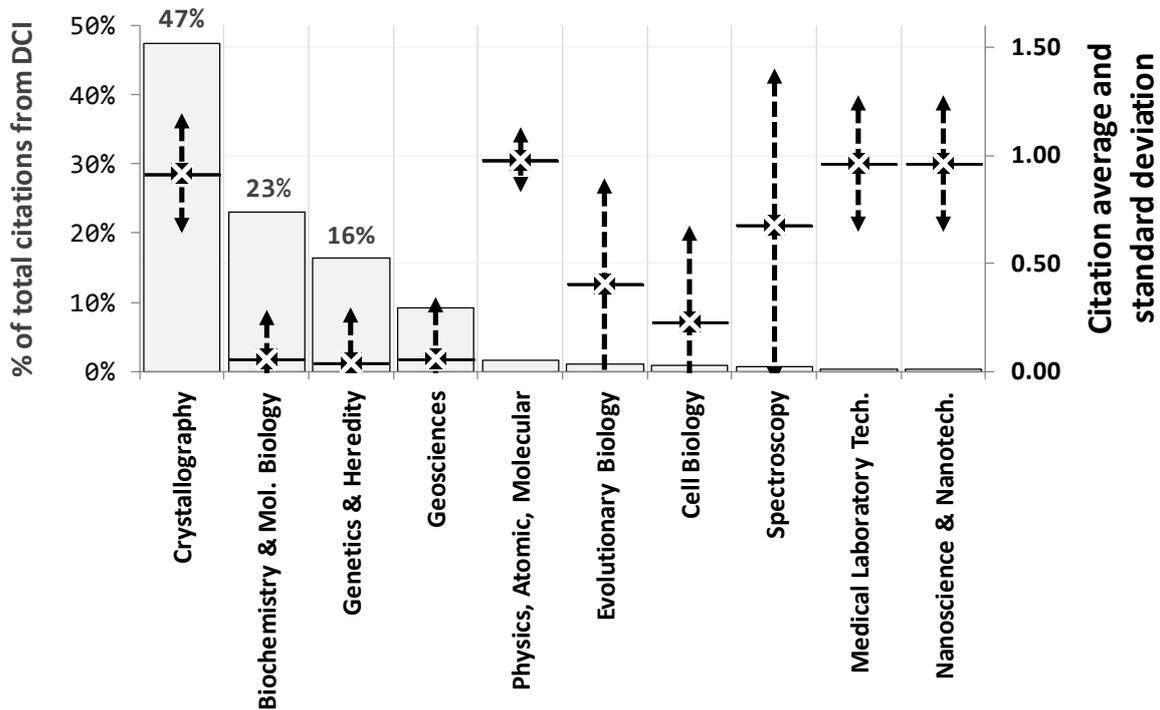

The pictures changes radically in the case of data studies (figure 2). Here, seven of the top ten categories belong to the area of Social Sciences. However, Biochemistry & Molecular Biology and Genetics & Heredity also make is to the top ten along with Health Care Sciences.

Figure 2. Top 10 subject categories with a higher number citations received, citation average and standard deviation for data studies indexed in the Data Citation Index.

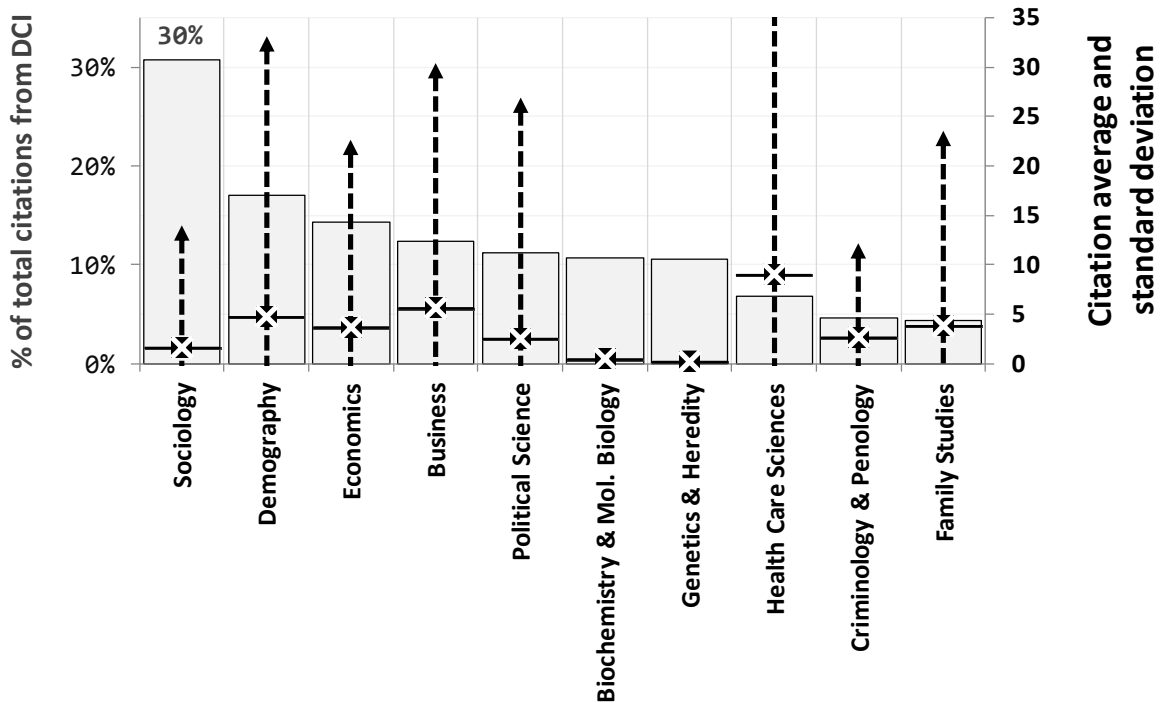

In order to explore if such accumulation of citations in specific categories is due to the in figure 3 we relate the number of records with the number of citations received for the largest repositories indexed in the DCI. Here, we see that the largest repository is specialized on Crystallography (Crystallography Open Database), followed by the Protein Data Bank and the Inter-university Consortium for Political and Social Research. Also, these three repositories are the ones containing a higher number of citations.

Figure 3. Main repositories in the DCI, citations received and total number of records

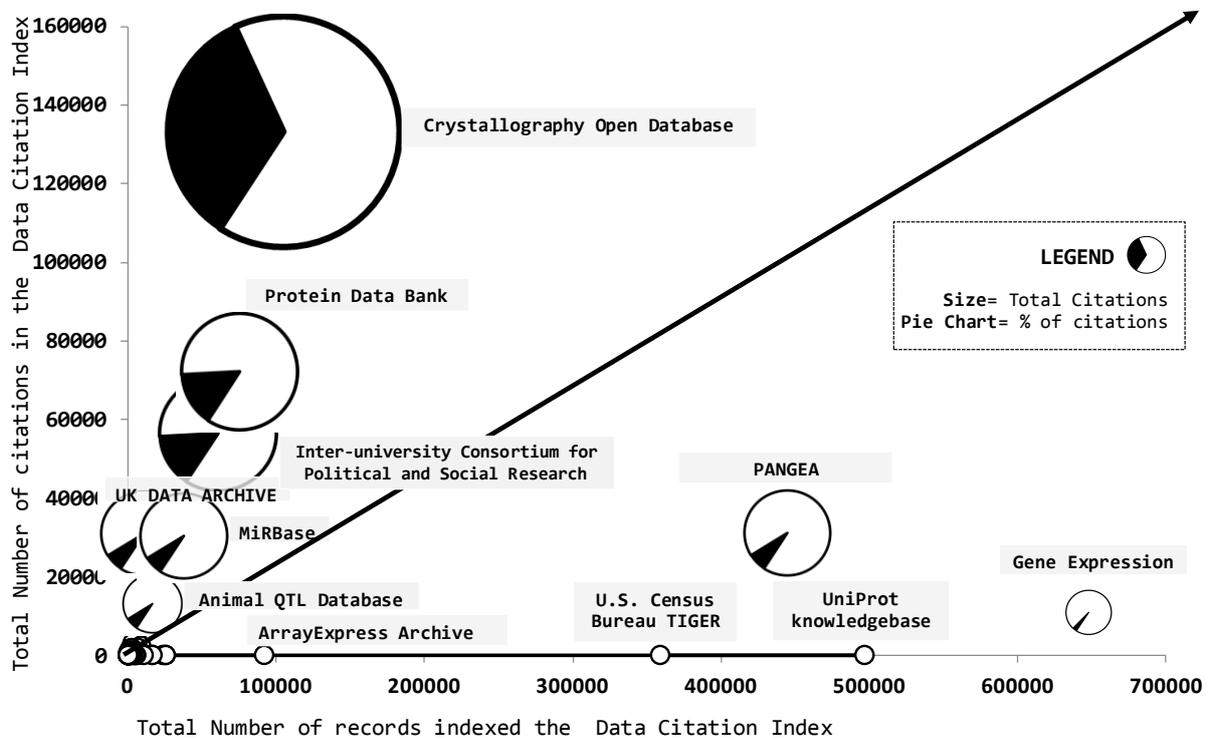

**Discussion and concluding remarks**

In this paper we conduct the first analysis on the citation distribution of the Thomson Reuters' Data Citation Index, a new database launched in 2012 which include a large number of data repositories associated with citation information. As observed, most of its records have no citations related with them, showing a high rate of uncitedness (88%). This demonstrates that data citation practices are far from common within the scientific community. Also, the DCI is heavily biased towards certain fields from the Hard Sciences as shown by Torres-Salinas, Martín-Martín & Fuente-Gutiérrez (2014), with almost no representation for Engineering & Technology which influences heavily the citation distribution. The reasons for this may not only be attributed to the criteria followed by Thomson Reuters, but to the expansion of data sharing practices within the research community. As indicated before, data sharing practices are not common to all areas of scientific knowledge and only certain fields have developed an infrastructure that allows to use and share data.

Even so, we observe different citation patterns depending on the area of study. While in Science and Engineering & Technology citations are concentrated among datasets, in the Social Sciences and Arts & Humanities, citations are normally referred to data studies. This fact is of extreme importance when conducting a citation analysis on data sharing practices as the chosen field will determine the suitability of focusing on one document type or the other. Similarly to what we see in scholarly publication.

The DCI seems a promising tool which may play an important role as data sharing expands among research fields. Citation analysis may encourage researchers to make their data publicly available as they will be able to analyze the impact of their contribution and the use of their work as well as developing a more open and transparent research process. In this sense, other repositories of a multidisciplinary nature have been launched in the recent years such as Figshare (http://figshare.com) which also seek at including metrics that will indicate the use and discussion wakened by the data displayed. Although data citation analyses do not seem yet appropriate as data sharing practices have not been fully expanded, the DCI seems to be a tool with a high potential in a near future.